\documentstyle[aps,preprint]{revtex}
\tightenlines
\begin{document}
\title{Gold nanotube: structure and melting} \draft
\author{G. Bilalbegovi\'c$^{*}$} \address{Department of Physics,
University of Rijeka, Omladinska 14, 51000 Rijeka,\\ Croatia}
\date{\today} \maketitle

\begin{abstract}
In the process of molecular dynamics simulation studies of gold nanowires an
interesting structure is discovered. This is a finite double-wall nanowire
with a large empty core similar to single-wall and double-wall carbon
nanotubes. The structure of the $16-10$ gold nanotube is studied at the room
temperature. An investigation of the high-temperature stability has also
been carried out. An unusual inward evaporation of atoms from cylindrical
liquid walls is found at $T \geq 1200$ K.


\vskip 20pt

$^{*}$This paper is the part of the invited talk ''Gold nanowires'' 
presented at JVC-9, June 16-20, 2002, Schloss Seggau by Graz, Austria,
to appear in Vacuum.

\end{abstract}

\section{Introduction}

Studies of nanoparticles are very important for advances in various fields
of technology. Carbon nanotubes are a topic of experimental and theoretical
research in an attractive and already well developed area of condensed
matter science \cite{CNT}. Single-wall and multi-wall carbon nanotubes, as
well as their bundles, are synthesized and their structural, thermal,
vibrational, mechanical, electronic and transport properties are
investigated. Cylindrical nanostructures made of other materials are also
the subject of research. For example, $Si$, $BN$, $SiSe_{2}$, $WS_{2}$, $%
MoS_{2}$, $NiCl_{2} $, and various metallic nanowires are studied \cite{CNT}.

Metallic nanowires are interesting from fundamental point of view. They are
also important for applications in nanomechanical and nanoelectronic
devices. Recently an extensive molecular dynamics (MD) study of unsupported
finite and infinite gold nanowires has been carried out \cite
{PRB,Fizika,MolSim,SSC,CompMatSci,JPCM,Tosatti,Baolin}. The systems analyzed
in simulations \cite{PRB,Fizika,MolSim,SSC,CompMatSci,JPCM} are in the range
of radii $R = (0.5 -1.5)$ nm, the lengths of finite nanowires are $L = (4 -
12)$ nm, and the numbers of particles are $N = (300 - 2100)$. It was found
that multi-wall nanowires of lasting stability often form. These gold
nanostructures consist of coaxial cylindrical sheets and resemble multi-wall
carbon nanotubes. Similar structures are imaged by transmission electron
microscopy (TEM) \cite{Kondo,Kizuka,Ugarte,Takai}. The string tension of
thinnest gold nanowires found in the TEM experiment \cite{Kondo} was
calculated by a MD simulation method \cite{Tosatti}. Changes of physical
properties in a transition from helical and multi-wall gold nanostructures
toward face-centered cubic (fcc) ones were analyzed by a genetic algorithm
and MD simulations \cite{Baolin}.

Multi-shell cylindrical nanostructures found in simulations \cite
{PRB,Fizika,MolSim,SSC,CompMatSci,JPCM} are preserved after a long
simulation time of $7$ ns. Nanowires whose initial diameters are larger than
their lenghts evolve toward an icosahedral shape, a well known structure in
cluster physics. Vibrational properties of several multi-wall nanowires are
investigated by diagonalization of the dynamical matrix \cite{PRB}. 
The averaged coordinates of particles from MD simulations are taken as an input.
The results show that the maximal frequencies calculated for cylindrical
multi-wall nanowires are higher than for the fcc bulk gold lattice. The
capacitance of finite nanometer-scale cylindrical capacitors is also
calculated and the values of the order of $0.5$ aF are found for length
scales where multi-wall nanowires appear in simulations \cite{Fizika}. Solid
multi-wall structures are stable up to rather high temperatures $T\sim 900$
K \cite{SSC}. Nanowires melt much below the bulk melting temperature of gold
($\sim 1350$ K for the potential used in simulations). Melting proceeds by
simultaneous disordering of all shells. Deformation properties of multi-wall
nanowires under axial compressive loading are studied at $T=300$ K \cite
{JPCM}. Several types of deformation are observed, for example large
buckling distortions and progressive crushing. It is found that compressed
nanowires recover their initial lengths and radii even after large
structural deformations. In contrast to the case of carbon nanotubes, in
gold nanowires irreversible local atomic rearrangements occur even under
small compressions. Multi-wall gold nanowires are able to sustain a large
compressive stress and to store mechanical energy. Structural and melting
properties of an unusual finite double-wall gold nanowire are described in
this work.

\section{Computational method}

In a MD simulation method the equations of motion for the system of
particles in a required configuration are solved numerically using suitable
algorithms. Simulations of gold nanowires are based on a well-tested
embedded atom potential \cite{Furio}. This potential has been shown to
accurately reproduce experimental values for a wide range of physical
properties of bulk gold, its surfaces and nanoparticles. A time step of $%
7.14\times 10^{-15}$ s is used. The temperature is controlled by rescaling
particle velocities. The MD box of a nanotube consists of 540 atoms. The
length of the box is $L=5$ nm, and the radius is $0.5$ nm. A finite nanowire
is constructed, i.e., the periodic boundary conditions are not used along
the wire axis. First, a nanowire with an ideal fcc structure and the
(111)-oriented cross section is prepared at $T=0$ K. This is done by
including all atoms whose distance from the wire axis is smaller than a
chosen radius. In previous simulations it was found that an (111) initial
orientation of cross sections produces better multi-wall structures than
(110) and (100) \cite{SSC}. An ideal sample is first relaxed at $T=0$ K.
Then, an annealing/quenching procedure is applied. The MD box is heated to $%
T=1000$ K. The resulting structure is a double-wall nanotube described in
this work. Cylindrical gold nanoparticles with the same radius and the
lengths of $2L$ and $3L$ are also simulated. Similar double-wall structures
with large empty cores are obtained.

\section{Results and discussion}

\subsection{Low-temperature structure}

A double-wall nanotube is shown in Fig. 1. The notation $n_1-n_2-n_3$...,
where $n_1>n_2>n_3$..., was introduced to label coaxial cylindrical shells
in multi-wall metallic nanowires \cite{Kondo}. Figure 1(a) presents a $16-10$
nanowire. As shown in Fig. 1(b), the nanotube is terminated by rounded caps
similar to capped carbon nanotubes. Figure 2 shows the central fragment of a
nanotube which proves that its walls are made of the triangular lattice of
gold atoms. The distance between the neighboring atoms is in the range $%
(0.25-0.29)$ nm. This should be compared with the bulk interatomic spacing
of fcc gold, $0.29$ nm. In TEM studies of suspended gold nanowires the
distance $d$ between dots on the images was measured \cite{Kondo}. These
dots represent positions of gold atoms projected on the image plane. It was
found that the distance between dots is in the range $d=(0.25-0.3)$ nm for
nanowires with diameters from $0.6$ nm to $1.5$ nm, and lengths from $3$ nm
to $15$ nm. The average value of this distance for $30$ nanowires is $d=0.288
$ nm \cite{Kondo}. These TEM studies of gold nanowires have shown that the
outer and inner tubes have the difference of the number of atom rows $%
n_1-n_2=7$. The exception is the finest studied nanowire $7-1$. However,
Kondo and Takayanagi have chosen to treat a single atom chain as "0". The
simulation presented here shows that a difference in the number of atom rows
is $6$, as in the nanowire $7-1$. A similarity between a simulated here $%
16-10$ nanowire and experimental $7-1$ is in their small radii.

\subsection{Melting}

The cross section of a nanotube at several temperatures is shown in Figs. 3
and 4. Close to the room temperature it is possible to distinguish atoms in
the trajectories plot. At higher temperatures atoms vibrate more strongly,
as shown in Fig. 3(a). At $T=900$ K diffusion in the walls is intensive, but
atoms do not move from the walls (see Fig. 3(b)). At $T=1200$ K, as shown in
Fig. 4(a), atoms in the liquid walls vibrate very strongly. Several atoms
evaporate into the empty core. Figure 4(b) shows that atoms evaporate into
the core, even when the walls begin to disarrange. Cylindrical liquid walls
exist before an evaporation starts. The opposite scenario of the
high-temperature disordering (not realized for multi-wall gold nanowires) is
a transition from a finite solid cylindrical structure to a solid and liquid
blob. The internal energy as a function of temperature is shown in Fig.
5(a). As in other simulated small nanowires \cite{SSC}, the jump in $E(T)$
as a clear sign of the first order phase transition in three-dimensional
systems is absent. Nanowires whose radii are of the order of $1$ nm are
close to one-dimensional systems for which the strict phase transitions do
not exist \cite{SSC}. The average mean-square displacement is shown in Fig.
5(b). The particle displacements sharply increase at $900$ K. Therefore, in
this nanotube melting starts much below the bulk melting temperature.

\section{Conclusions}

Recently an evidence of a single-wall platinum nanotube was reported \cite
{Oshima}. This structure was obtained from a suspended platinum nanowire by
electron-beam thinning method and was imaged in ultrahigh-resolution
electron microscope. The outer shell of a $13-6$ double-wall platinum
nanowire was stripped and inner shell was exposed. The results presented
here show the existence of an unsupported double-wall gold nanotube.

Gold nanowires are very promising for applications in nanodevices, for
example as interconnections in integrated circuits. 
Experimental results \cite{Kondo,Kizuka,Ugarte,Takai}, as well as
simulations presented
here and in Refs. \cite{PRB,Fizika,MolSim,SSC,CompMatSci,JPCM,Tosatti,Baolin},
show that the smallest gold nanowires are strong and stable even at high
temperatures. These nanowires exist in various exotic forms, from monatomic
chains to multi-wall cylindrical structures \cite
{PRB,Fizika,MolSim,SSC,CompMatSci,JPCM,Tosatti,Baolin,Kondo,Kizuka,Ugarte,Takai}%
. Transport properties of gold nanowires may also make them very useful \cite
{Friedrich}. In preparation and investigation of metallic nanowires results
obtained by molecular dynamics simulations are important.

\acknowledgments

This work has been carried under the HR-MZT project 119206 ``Dynamical
Properties of Surfaces and Nanostructures'' and the EC Research Action COST
P3 ``Simulation of Physical Phenomena in Technological Applications''.

\clearpage

\begin{figure}[tbp]
\caption{Atomic positions for a nanotube at $T=300$ K: (a) top view of a
cross section after the caps were removed. This is a $16-10$ nanowire (atoms
from two layers are completely visible), (b) side view.}
\label{fig1}
\end{figure}

\begin{figure}[tbp]
\caption{Atomic positions for a side view of partial walls at T=300 K. An
irregular cut of a nanotube containing atoms from both walls is presented to
show a cylindrical triangular lattice.}
\label{fig2}
\end{figure}

\begin{figure}[tbp]
\caption{The structural properties of a nanotube: (a) $T=500$ K, (b) $T=900$
K. These top views of the central part of a nanotube are represented by the
particle trajectory plots and refer to a time span of $3.5$ ps. All atoms in
the central slice of the thickness of $4$ nm along the nanotube axis are
included.}
\label{fig3}
\end{figure}

\begin{figure}[tbp]
\caption{The high-temperature properties of a nanotube: (a) $T=1200$ K, (b) $%
T=1300$ K. Details as in Fig. 3.}
\label{fig4}
\end{figure}

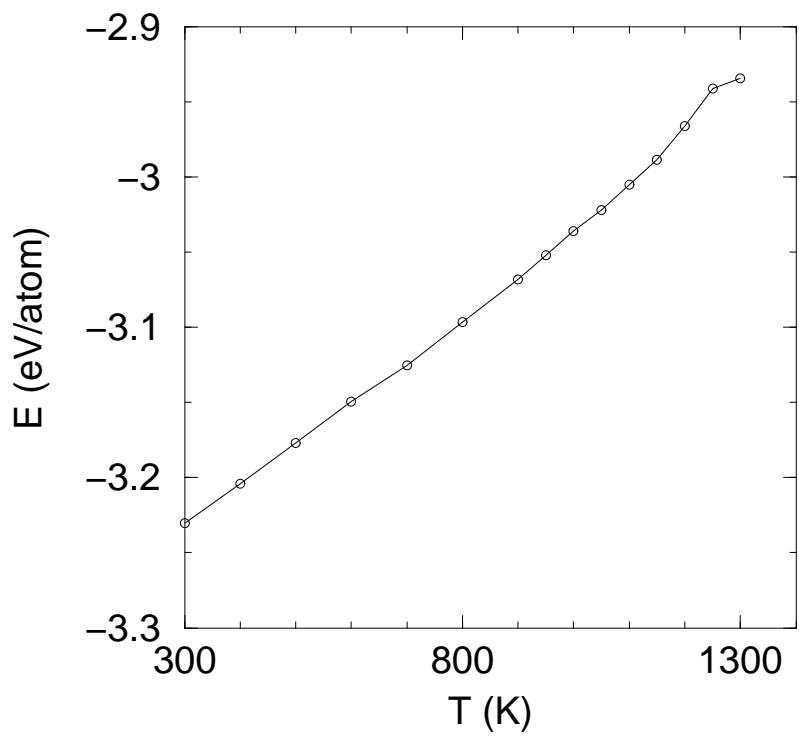
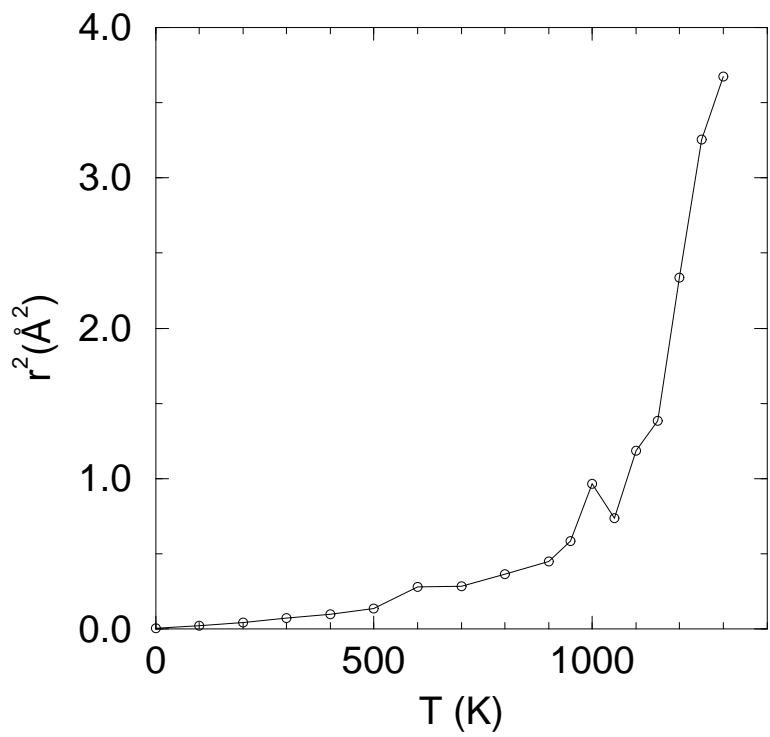
\begin{figure}[tbp]
\caption{ The temperature dependence of: (a) total energy per atom, (b)
mean-square displacements.}
\label{fig5}
\end{figure}

\end{document}